\documentclass[12pt]{article}
\usepackage{float} 
\usepackage{amsmath}
\usepackage{slashed}
\usepackage{amsfonts}   
\usepackage{amssymb}

\begin{document}
\date{}
\title{{\bf{\Large Semiclassical dynamics for torsional  Newton-Cartan strings}}}
\author{
 {\bf {\normalsize Dibakar Roychowdhury}$
$\thanks{E-mail:  dibakarphys@gmail.com, dibakarfph@iitr.ac.in}}\\
 {\normalsize  Department of Physics, Indian Institute of Technology Roorkee,}\\
  {\normalsize Roorkee 247667, Uttarakhand, India}
\\[0.3cm]
}

\maketitle
\begin{abstract}
We explore \emph{folded} spinning string configurations over torsional Newton Cartan (TNC) geometry with $ R\times S^2 $ topology within the semiclassical approximation. We consider the large $ c $ and/or nonrelativistic (NR) limit associated with the world-sheet d.o.f.  and compute the one loop stringy corrections to the energy spectrum in the dual Spin Matrix Theory (SMT) theory in the limit of strong ($ \mathtt{g}\gg 1 $) coupling.
\end{abstract}
\section{Overview and Motivation}
The quest for a consistent (UV finite) low energy description of relativistic string theory had been one of the active areas of theoretical investigations for last couple of decades \cite{Gomis:2000bd}-\cite{Gomis:2005pg}. Apparently there seems to be existing two parallel formulations of nonrelativistic (NR) string theories over curved target space geometries. One of these formulations is based on taking $ 1/c $ limit of General Relativity (GR) in the first order formalism that eventually leads towards curved manifold structures (known as \emph{string Newton-Cartan} (SNC) geometry) together with a flux-less auxiliary two form field \cite{Andringa:2012uz}-\cite{Bergshoeff:2019pij}. It turns out that under such circumstances one could in fact define a NR quantum consistent 2D sigma model that is invariant under transformations generated by SNC generators \cite{Bergshoeff:2018yvt}-\cite{Gomis:2019zyu}. In other words, NR closed string spectra on SNC background may be obtained through large $ c $ expansion of the relativistic Nambu-Goto (NG) action.

The other approach is based on the formulation of 2D sigma models on curved manifolds called torsional Newton-Cartan (TNC) geometries \cite{Harmark:2017rpg}-\cite{Roychowdhury:2020kma}.
TNC strings are obtained through target space null reduction of Poincare invariant 2D string sigma models while keeping the string momentum along the null isometry direction fixed\footnote{Recently, the authors in \cite{Harmark:2019upf} have shown that under certain specific assumptions on the compact longitudinal spatial direction, the TNC strings (in the presence of background NS-NS fluxes) can be mapped to NR strings propagating over SNC geometry. }. Taking a zero tension limit of TNC strings finally leads towards an emerging new sector in the celebrated gauge/string duality. For example, in the case of $ AdS_5 \times S^5 $ (super)strings such scaling results in the so called NR 2D sigma model/Spin Matrix Theory (SMT) correspondence \cite{Harmark:2017rpg}-\cite{Harmark:2018cdl}. The NR 2D sigma model thus obtained has been found to possess an underlying Galilean Confromal Algebra together with a NR Weyl invariance. On the other hand, in the case of $ AdS_5 \times S^5 $ strings, the SMT limit corresponds to taking a NR limit of the relativistic magnon dispersion relation in $ \mathcal{N}=4 $ SYM \cite{Harmark:2008gm}-\cite{Harmark:2014mpa}.

The central idea that lives at the heart of the gauge/string duality is the identification of the energy spectra ($ \Delta $) (associated with gauge invariant operators in the dual gauge theory) to that with the stringy excitations ($ E $) over curved background geometry. In other words, the duality conjecture drives us towards a natural identification of \emph{heavy} operators and/or gauge theory states with solitonic excitations associated to 2D sigma models in the supergravity approximations. The first paper to address this issue is due to BMN \cite{Berenstein:2002jq} who showed that semiclassical type IIB string states on ``pp wave background" could be estimated \emph{exactly} and therefore compared with the corresponding spectrum associated with \emph{heavy} states in the dual $ \mathcal{N}=4 $ SYM in the non BPS sector at strong coupling. This idea was further pushed forward due to authors in \cite{Gubser:2002tv} who constructed semiclassical states due to spinning ($ S\gg 1 $) closed string configurations in $ AdS_5\times S^5 $ and in particular obtained so called cusp anomalous dimension corresponding to operators in the dual $ \mathcal{N}=4 $ SYM at strong coupling. Subsequently, several generalizations of these ideas came forward \cite{Frolov:2002av}-\cite{Arutyunov:2003za}.

So far these questions have been addressed mostly in the context of relativistic string sigma model/CFT correspondence as mentioned above. An answer to these issues in the context of NR sigma model/gauge theory correspondence is still lacking in the literature. The purpose of the present analysis is therefore to pursue similar questions in the context of NR strings/SMT correspondence \cite{Harmark:2017rpg} where (considering certain specific limits \cite{Harmark:2014mpa}) we explore gauge theory states (with large quantum numbers) in the strongly coupled SMT sector by probing solitonic (stringy) excitations associated to NR 2D sigma model on $ R\times S^2 $.  The approach of the present paper might be considered as a NR counterpart of \cite{Gubser:2002tv} which we elaborate in detail in the following paragraph below.

The set of gauge invariant operators (in the dual SMT theory at strong coupling and low temperatures) that we choose to work with are those with large R charge ($ \tilde{\mathtt{J}}_{\varphi} \gg1$) and large energy quantum numbers ($ \tilde{\Delta}_{NR}\gg 1 $). On the string theory side, we realize these states as solitonic excitations \cite{Harmark:2014mpa} associated to closed NR folded spinning string configurations over $ R\times S^2 $. These strings are supposed to be stretched along the polar coordinate of $ S^2 $ and spinning around their center of mass that is happened to be coincident with the north pole of $ S^2 $ \cite{Gubser:2002tv}.  A straightforward computation on the stringy side reveals (\ref{e87}),
\begin{eqnarray}
\tilde{\Delta}_{NR}-\tilde{\mathtt{J}}_{\varphi}=\tilde{\tilde{\mathtt{f}}}(\bar{\mathtt{g}})\tilde{\mathtt{J}}_{\varphi}
\label{anomalous}
\end{eqnarray}
where we compute the function,
\begin{eqnarray}
\tilde{\tilde{\mathtt{f}}}(\bar{\mathtt{g}})=\mathfrak{q}_1\bar{\mathtt{g}}\left( 1+\frac{\mathfrak{q}_2}{\sqrt{\tilde{\mathtt{J}}_{\varphi}}}\frac{1}{\bar{\mathtt{g}}^{1/4}}+..\right) 
\end{eqnarray}
upto one loop in the string corrections. Here, $ \bar{\mathtt{g}}(=\frac{\mathtt{g}}{\tilde{\mathtt{J}}_{\varphi}^{2}} )\ll 1$ should be regarded as being the \emph{effective} expansion parameter in the dual SMT in the regime of strong $ (\mathtt{g}=c^2 \lambda\gg1 )$ coupling where, $ \lambda \ll 1 $ is the standard t'Hooft coupling in the $ \mathcal{N}=4 $ SYM theory and $ c(\rightarrow \infty) $ is the speed of light \cite{Harmark:2017rpg}-\cite{Harmark:2018cdl}. The above result (\ref{anomalous}) is therefore a non-perturbative effect from the perspective of the dual SMT and is the key finding of the present analysis.

\section{NR spinning strings on $ R\times S^2 $}
Our analysis starts with a formal construction of the spinning string sigma model over torsional Newton-Cartan (TNC) geometry \cite{Harmark:2017rpg}-\cite{Harmark:2018cdl} with $ R \times S^2 $ topology \cite{Grosvenor:2017dfs},
\begin{eqnarray}
ds_{TNC}^2 = 2 \tau (d \mathfrak{u}-\mathfrak{m})+\mathfrak{h}_{\mu \nu}dx^{\mu}dx^{\nu}\label{e1}
\end{eqnarray}
where each of the individual entities could be formally expressed as \cite{Roychowdhury:2019olt},
\begin{eqnarray}
\tau &=&\tau_{\mu}dx^{\mu}=\frac{1}{2}d\psi +dt-\frac{1}{2}\cos\theta d\varphi  ~;~ \mathfrak{u}=\frac{\psi}{4}-\frac{t}{2}\nonumber\\
\mathfrak{m}&=&\mathfrak{m}_{\mu}dx^{\mu} =\frac{1}{4}\cos\theta d\varphi ~;~
\mathfrak{h}_{\mu \nu}dx^{\mu}dx^{\nu} =\frac{1}{4}\left[ d\theta^{2}+\sin^2 \theta d\varphi^{2}\right].\label{e2}
\end{eqnarray}

Notice that, here $ \mathfrak{u} $ and $ \varphi $ are the \textit{isometry} directions associated with the target space geometry \cite{Grosvenor:2017dfs},\cite{Roychowdhury:2019olt}. The relativistic action corresponding to the \emph{bosonic} sector of the 2D string sigma model could be formally expressed as\footnote{Here $ L $ is the radius of the 2 sphere.}, 
\begin{eqnarray}
\mathcal{S}_{NG} =\int d^2\sigma \mathcal{L}_{NG}=- \frac{\sqrt{\lambda}}{4 \pi}\int d\tau d\sigma \tilde{\mathcal{L}}_{NG}~;~\sqrt{\lambda}=\frac{L^2}{\alpha'}
\label{e4}
\end{eqnarray}
where we identify the corresponding sigma model Lagrangian as\footnote{This is the Lagrangian that represents \emph{relativistic} strings propagating over TNC geometry in the semiclassical approximation. The world-sheet theory is nonrelativisitc (NR) only from the perspective of the target space geometry given in the problem \cite{Harmark:2017rpg}-\cite{Harmark:2018cdl}. However, it is always possible to consider a large $ c(\rightarrow \infty) $ limit \cite{Harmark:2017rpg}-\cite{Harmark:2018cdl} associated to 2D sigma models that eventually leads towards NR strings propagating over nonrelativistic $ U(1) $ Galilean geometry. } \cite{Harmark:2018cdl},\cite{Roychowdhury:2019olt}
\begin{eqnarray}
\tilde{\mathcal{L}}_{NG}=\frac{\varepsilon^{\alpha \alpha'}\varepsilon^{\beta \beta'}}{\varepsilon^{\alpha \alpha'}\chi_{\alpha}\partial_{\alpha'}\zeta}(\partial_{\alpha'}\zeta \partial_{\beta'}\zeta -\chi_{\alpha'}\chi_{\beta'})\left( \partial_{\alpha}\theta \partial_{\beta}\theta +\sin^{2}\theta\partial_{\alpha}\varphi \partial_{\beta}\varphi\right)-\varepsilon^{\alpha \beta}\cos\theta \partial_{\alpha}\varphi \partial_{\beta}\zeta.
\label{E4}
\end{eqnarray}

Notice that, here $ \zeta $ is the additional compact direction associated with the target space geometry along which the string has non zero windings \cite{Harmark:2017rpg}-\cite{Harmark:2018cdl}. Moreover, here $ \varepsilon^{01}=-\varepsilon_{01}=1 $ is the Levi-Civita symbol in 2D together with\footnote{Here $ \chi_{\alpha} $ essentially is the pullback of the Newton-Cartan clock one form $ \tau_{\mu} $\cite{Harmark:2017rpg} namely, $ \chi_{\alpha} \equiv \tau_{\alpha}=\tau_{\mu}\partial_{\alpha}X^{\mu}. $} \cite{Roychowdhury:2019olt},
\begin{eqnarray}
\chi_{\alpha}=2\partial_{\alpha}t+\partial_{\alpha}\psi - \cos\theta \partial_{\alpha}\varphi.
\end{eqnarray}

The goal of the present analysis is to find out the Hamiltonian spectrum corresponding to NR folded strings spinning configurations over $ U(1) $ Galilean geometry with $ R \times S^2 $ topology. This is achieved by taking the so called \emph{scaling limit} \cite{Harmark:2017rpg},\cite{Roychowdhury:2019olt} of the relativistic sigma model Lagrangian\footnote{Here, $ \mathtt{g} (=c^2 \lambda)$ is the string tension associated to TNC strings in the large $ c(\rightarrow \infty) $ limit.} (\ref{E4}),
\begin{eqnarray}
\lambda = \frac{\mathsf{g}}{c^2}~;~t= c^{2}\tilde{t}~;~\psi =\tilde{\psi}~;~\theta = \tilde{\theta}~;~\varphi = \tilde{\varphi}~;~\zeta = c~ \tilde{\zeta}.
\end{eqnarray}

The resulting NR sigma model action could be formally expressed as,
\begin{eqnarray}
\mathcal{S}_{NR}=\frac{\sqrt{\mathtt{g}}}{4\pi}\int d^2\sigma \mathcal{L}_{NR}
\end{eqnarray}
together with the identification of the NR Lagrangian density as \cite{Roychowdhury:2019olt},
\begin{eqnarray}
 \mathcal{L}_{NR}=\frac{\varepsilon^{\alpha \alpha'}\varepsilon^{\beta \beta'}}{\dot{t}}\partial_{\alpha'}t\partial_{\beta'}t(\partial_{\alpha}\theta \partial_{\beta}\theta + \sin^2\theta \partial_{\alpha}\varphi \partial_{\beta}\varphi)+\varepsilon^{\alpha \beta}\cos\theta \partial_{\alpha}\varphi \partial_{\beta}\zeta +\mathcal{O}(1/c^2)
 \label{e66}
\end{eqnarray}
where in order to simplify calculations we remove tildes on wards.
\section{Classical solutions}
We start with the folded spinning string ansatz of the following form,
\begin{eqnarray}
t=\tilde{\kappa}\tau ~;~ \varphi =\tilde{\varpi} \tau ~;~\zeta =\tilde{\ell} \sigma ~;~\theta = \theta (\sigma).
\label{e67}
\end{eqnarray}

Substituting (\ref{e67}) into (\ref{e66}) we find,
\begin{eqnarray}
 \mathcal{L}_{NR}=\tilde{\kappa}\theta'^2 +\tilde{\varpi}\tilde{\ell}\cos\theta.
\end{eqnarray}

The corresponding equation of motion turns out to be,
\begin{eqnarray}
\theta''(\sigma) +\frac{\tilde{\varpi}\tilde{\ell}}{2\tilde{\kappa}}\sin\theta = 0
\label{e69}
\end{eqnarray}
which has the most general solution of the following form\footnote{The entity introduced on the r.h.s. of (\ref{e70}) is known as the \emph{Jacobi amplitude} ($ \phi(u,k)\equiv  \text{am}(u|k)=\int_{0}^{u} dn(u',k)du'$ where, $ dn(u',k)du' $ is known as \emph{Jacobi elliptic function} with elliptic modulus) which is defined as being the inverse of the elliptic integral function of first kind. For more details on different types of Jacobi elliptic functions the enthusiastic reader is encouraged to see \cite{abramh}.},
\begin{eqnarray}
\theta (\sigma)\equiv \theta_{cl}=2 \text{am}\left(\frac{\sqrt{\tilde{\varpi}\tilde{\ell}+\tilde{\kappa} c_1} \left(\sigma +c_2\right)}{2 \sqrt{\tilde{\kappa}}}|\frac{2 \tilde{\varpi}\tilde{\ell}}{\tilde{\varpi}\tilde{\ell}+\tilde{\kappa} c_1}\right).
\label{e70}
\end{eqnarray}

Next, we note down the corresponding conserved charges namely the energy ($ \tilde{\mathcal{E}} $) and the R charge ($ \tilde{\mathtt{J}}_{\varphi} $) associated with the stringy dynamics,
\begin{eqnarray}
\tilde{\mathcal{E}}&=&\frac{\sqrt{\mathtt{g}}}{4\pi}\int_{0}^{2\pi} \theta'^2 d\sigma\nonumber\\
 &=&\frac{\sqrt{\mathtt{g}}}{\pi} \left(\text{am}\left(\frac{\sqrt{\tilde{\varpi}\tilde{\ell}+\tilde{\kappa} c_1} \left(c_2+2 \pi \right)}{2 \sqrt{\tilde{\kappa}}}|\frac{2 \tilde{\varpi}\tilde{\ell}}{\tilde{\varpi}\tilde{\ell}+\tilde{\kappa} c_1}\right)^2-\text{am}\left(\frac{\sqrt{\tilde{\varpi}\tilde{\ell}+\tilde{\kappa} c_1} c_2}{2 \sqrt{\tilde{\kappa}}}|\frac{2 \tilde{\varpi}\tilde{\ell}}{\tilde{\varpi}\tilde{\ell}+\tilde{\kappa} c_1}\right)^2\right)\nonumber\\
 &\equiv &\frac{\sqrt{\mathtt{g}}}{\pi}~ \mathtt{E}
 \label{e71}
\end{eqnarray}
and\footnote{See Appendix for details.},
\begin{eqnarray}
\tilde{\mathtt{J}}_{\varphi}&=&\frac{\sqrt{\mathtt{g}}\tilde{\ell}}{4\pi}\int_{0}^{2\pi} \cos\theta d\sigma =\frac{\sqrt{\mathtt{g}}\tilde{\ell}}{4\pi}\int_{0}^{2\pi } \sqrt{1-\frac{4\tilde{\kappa}^2}{\tilde{\varpi}^2 \tilde{\ell}^2}(\theta'')^2}d\sigma \nonumber\\
 &=&\frac{\sqrt{\mathtt{g}}\tilde{\ell}\sqrt{\tilde{\kappa}}}{4\pi}\frac{\mathtt{N}}{\mathtt{D}}.
 \label{e72}
\end{eqnarray}

Using (\ref{e71}) and (\ref{e72}) it is now straightforward to obtain\footnote{Here $ \tilde{\Delta}_{NR} $ is the energy eigenvalue associated with the corresponding dual operator spectrum in the SMT theory at strong coupling and low temperatures.},
\begin{eqnarray}
\tilde{\Delta}_{NR}\sim \tilde{\mathcal{E}}=\tilde{\mathtt{J}}_{\varphi}\left(1+\tilde{\mathtt{f}}(\bar{\mathtt{g}}) \right) 
\end{eqnarray}
where we identify,
\begin{eqnarray}
\tilde{\mathtt{f}}(\bar{\mathtt{g}}) =\frac{\bar{\mathtt{g}}}{4\pi^2}\frac{\tilde{\ell}\sqrt{\tilde{\kappa}}\mathtt{N}}{\mathtt{D}}\left( \mathtt{E}-\frac{\tilde{\ell}\sqrt{\tilde{\kappa}}}{4}\frac{\mathtt{N}}{\mathtt{D}}\right)\equiv \mathfrak{q}_{1} \bar{\mathtt{g}}
\label{e74}
\end{eqnarray}
as being the leading order correction to the Hamiltonian spectrum in the limit of strong ($ \mathtt{g}\gg 1 $) coupling. Here we identify, $ \bar{\mathtt{g}}=\frac{\mathtt{g}}{\tilde{\mathtt{J}}^{2}_{\varphi}} $ as an effective coupling constant in the dual SMT theory at low temperatures.
\section{One loop corrections}
Our aim now is to compute quantum $(\sqrt{\mathtt{g}} )^{-1} $ corrections to the above function in (\ref{e74}). In order to do so, we choose to work with the string embedding of the following form,
\begin{eqnarray}
t=\tilde{\kappa}\tau +\frac{\hat{t}(\sigma^{\alpha})}{\mathtt{g}^{1/4}}~;~ \varphi =\tilde{\varpi} \tau +\frac{\hat{\varphi}(\sigma^{\alpha})}{\mathtt{g}^{1/4}} ~;~\zeta =\tilde{\ell} \sigma +\frac{\hat{\zeta}(\sigma^{\alpha})}{\mathtt{g}^{1/4}}~;~\theta =\theta_{cl} (\sigma)+\frac{\hat{\theta} (\sigma)}{\mathtt{g}^{1/4}}.
\label{e75}
\end{eqnarray}

Substituting (\ref{e75}) into (\ref{e66}) we arrive at the quadratic Lagrangian of the following form,
\begin{eqnarray}
\mathcal{L}^{(2)}_{NR}=\tilde{\kappa}\hat{\theta}'^2 +2\hat{\theta}' \theta_{cl}'+\mathfrak{w}\mathfrak{l}\cos\theta_{cl}-(\tilde{\varpi}\mathfrak{l}+\tilde{\ell}\mathfrak{w})\hat{\theta}\sin\theta_{cl}
\end{eqnarray}
where we choose to work with the following ansatz, 
\begin{eqnarray}
\hat{t}(\sigma^{\alpha})=\tau ~;~\hat{\varphi}(\sigma^{\alpha})=\mathfrak{w}\tau ~;~\hat{\zeta}(\sigma^{\alpha})=\mathfrak{l}\sigma 
\end{eqnarray}
together with the fact that $ \mathfrak{w} $ and $ \mathfrak{l} $ are some real positive integers.

The equation of motion corresponding to $ \hat{\theta}(\sigma) $ could be formally expressed as,
\begin{eqnarray}
2\tilde{\kappa}\hat{\theta}''(\sigma)+2\theta_{cl}''(\sigma)+(\tilde{\varpi}\mathfrak{l}+\tilde{\ell}\mathfrak{w})\sin\theta_{cl}=0.
\label{e78}
\end{eqnarray}

The above equation (\ref{e78}) has a remarkably simple solution once we set,
\begin{eqnarray}
\tilde{\kappa}=\tilde{\varpi}=\tilde{\ell}=\mathfrak{w}=\mathfrak{l}=1
\end{eqnarray}
which by means of (\ref{e69}) yields,
\begin{eqnarray}
\hat{\theta}(\sigma)=\theta_{cl}(\sigma)+\mathfrak{s}_1\sigma +\mathfrak{s}_2
\label{e80}
\end{eqnarray}
where $ \mathfrak{s}_{1,2} $ are integration constants.

Using (\ref{e80}), it is now straightforward to compute one loop stringy correction to the energy spectrum\footnote{For the sake of computational simplicity we set, $ c_1=c_2=\mathfrak{s}_1=1 $ and $ \mathfrak{s_2}=0 $.},
\begin{eqnarray}
\tilde{\mathcal{E}}=\frac{\sqrt{\mathtt{g}}}{\pi}~\left(  \mathtt{E}+\frac{\sqrt{\hat{\alpha}'}}{L}\Delta \mathtt{E}\right) 
\label{e81}
\end{eqnarray}
where we identify one loop correction to the NR string spectrum as\footnote{Notice that here, $ \text{gd}(x)=\int_{0}^{x}\frac{dt}{\cosh t} $ is the so called \emph{Gudermannian} function \cite{abramh}. },
\begin{eqnarray}
 \Delta \mathtt{E}=-\text{gd}\left(\frac{1}{\sqrt{2}}\right)+\text{gd}\left(\frac{1+2 \pi }{\sqrt{2}}\right)+\sqrt{2} \sinh \left(\sqrt{2} \pi \right) \text{sech}\left(\frac{1}{\sqrt{2}}\right) \text{sech}\left(\frac{1+2 \pi }{\sqrt{2}}\right).
\end{eqnarray} 

Finally, we compute one loop correction to the R charge,
\begin{eqnarray}
\tilde{\mathtt{J}}_{\varphi}=\frac{\sqrt{\mathtt{g}}}{4\pi}~\left( \frac{\mathtt{N}}{\mathtt{D}}+\frac{\sqrt{\hat{\alpha}'}}{L}\Delta \mathtt{J}\right) \label{e83}
\end{eqnarray} 
where,
\begin{eqnarray}
\Delta \mathtt{J}=\frac{4 \sqrt{2} \left(2 e^{\frac{1+2 \pi }{\sqrt{2}}} \left(\text{gd}\left(\frac{1+2 \pi }{\sqrt{2}}\right)+\pi \right)+3\right)}{1+e^{\sqrt{2} (1+2 \pi )}}
-\frac{4 \sqrt{2} \left(2 e^{\frac{1}{\sqrt{2}}} \text{gd}\left(\frac{1}{\sqrt{2}}\right)+1\right)}{1+e^{\sqrt{2}}}\nonumber\\
-4 \sqrt{2}+2 \pi +6+8 \tan ^{-1}\left(e^{\frac{1}{\sqrt{2}}}\right)-8 \tan ^{-1}\left(e^{\frac{1+2 \pi }{\sqrt{2}}}\right)-\sqrt{2} \cosh ^{-1}(3).
\end{eqnarray}

Combining (\ref{e81}) and (\ref{e83}) we finally obtain,
\begin{eqnarray}
\tilde{\Delta}_{NR}\sim \tilde{\mathcal{E}}=\tilde{\mathtt{J}}_{\varphi}(1+\tilde{\tilde{\mathtt{f}}}(\bar{\mathtt{g}}))
\label{e87}
\end{eqnarray}
where we identify the function,
\begin{eqnarray}
\tilde{\tilde{\mathtt{f}}}(\bar{\mathtt{g}})=\tilde{\mathtt{f}}(\bar{\mathtt{g}})(1+\Delta \tilde{\mathtt{f}})
\end{eqnarray}
that include one loop stringy correction of the form,
\begin{eqnarray}
\Delta \tilde{\mathtt{f}}&=&\frac{\sqrt{\hat{\alpha}'}}{L}\left(\frac{\Delta \mathtt{E}+\Delta \mathtt{S}\frac{\mathtt{D}}{\mathtt{N}}\left(\mathtt{E}-\frac{\mathtt{N}}{2\mathtt{D}} \right) }{\mathtt{E}-\frac{\mathtt{N}}{4\mathtt{D}}  } \right) +\mathcal{O}(\hat{\alpha}'/L^2)\nonumber\\
&=&\frac{\mathfrak{q}_2}{\sqrt{\tilde{\mathtt{J}}_{\varphi}}}\frac{1}{\bar{\mathtt{g}}^{1/4}}+\mathcal{O}(1/(\tilde{\mathtt{J}}_{\varphi}\sqrt{\bar{\mathtt{g}}}))
\label{e89}
\end{eqnarray}
where, $ \mathfrak{q}_{2}=\left(\frac{\Delta \mathtt{E}+\Delta \mathtt{S}\frac{\mathtt{D}}{\mathtt{N}}\left(\mathtt{E}-\frac{\mathtt{N}}{2\mathtt{D}} \right) }{\mathtt{E}-\frac{\mathtt{N}}{4\mathtt{D}}  } \right) $.

Finally, substituting (\ref{e89}) into (\ref{e87}) we find,
\begin{eqnarray}
\tilde{\Delta}_{NR}=\left( 1+\mathfrak{q}_1\bar{\mathtt{g}}+\frac{\mathfrak{q}_1\mathfrak{q}_2}{\sqrt{\tilde{\mathtt{J}}_{\varphi}}}\bar{\mathtt{g}}^{3/4}+..\right) \tilde{\mathtt{J}}_{\varphi}.
\label{e90}
\end{eqnarray}
\section{Summary and final remarks}
The present paper is an attempt towards understanding the NR sigma model/SMT correspondence through some explicit computations of the energy spectrum in the dual gauge theory at strong ($ \mathtt{g}\gg 1 $) coupling. In our analysis, we consider specific example of NR folded closed string configurations on $ R\times S^2 $ spinning around the north pole of $ S^2 $.  By exploring solitonic excitations associated with these NR 2D sigma models we finally compute the Hamiltonian spectrum ($ \tilde{\Delta}_{NR} $) corresponding to \emph{heavy} states ($ \mathtt{\tilde{J}}_{\varphi}>\sqrt{\mathtt{g}}\gg 1 $) in the dual gauge theory at strong coupling. In other words, on the gauge theory side we consider states with large length ($ L \sim  \mathtt{\tilde{J}}_{\varphi}$) which thereby implies an effective planar limit at low temperatures. On the dual stringy sector this limit corresponds to \emph{free} strings propagating over $ S^2 $ \cite{Harmark:2014mpa}. The key observation of our analysis turns out to be the identification of an effective expansion parameter $ \bar{\mathtt{g}}(=\frac{\mathtt{g}}{\tilde{\mathtt{J}}_{\varphi}^{2}} )\ll 1$  for the gauge theory which allows us to compute quantum corrections associated with the Hamiltonian spectra (around its classical value) in the limit of strong ($ \mathtt{g}\gg 1 $) coupling and low temperatures. An interesting question along this line of arguments turns out to be how does the spectrum changes (at strong coupling) as temperature increases above the critical transition temperature ($ T_c $) where the classical stringy picture breaks down and one therefore needs to take into account the so called \emph{non planar} effects \cite{Harmark:2014mpa}. The other question that might be of worth exploring is whether it is possible to match the spectrum at \textit{finite} (or, \emph{small}) coupling ($ \mathtt{g} $). In other words, to check the spin chain (or, near planar $ \mathcal{N}=4 $ SYM)/ NR sigma model correspondence using the notion of effective coupling constant ($\bar{\mathtt{g}}  $) discussed in this paper. We hope to address some of these questions in the near future.\\ \\ 
{\bf {Acknowledgements :}}
 The author is indebted to the authorities of IIT Roorkee for their unconditional support towards researches in basic sciences. \\\\\\\\\\\\\\\\\\\\\\\\\\\\
{ \large{ {\bf {Appendix: Detailed expressions for $\mathtt{N}$ and $ \mathtt{D} $}}}}\\ 
Here, in the Appendix, we provide detailed expressions for the functions $\mathtt{N}$ and $ \mathtt{D} $ those which appear in the expression for the R charge (\ref{e72}) associated to NR TNC strings spinning over $ R\times S^2 $. Below we enumerate their individual expressions as, 
\begin{eqnarray*}
\mathtt{N}=-\left(c_1 c_2 \sqrt{ \tilde{\kappa}} \sqrt{c_1  \tilde{\kappa}+\tilde{\varpi}\tilde{\ell}}-2 \left(c_1  \tilde{\kappa}+\tilde{\varpi}\tilde{\ell}\right) E\left(\text{am}\left(\frac{\sqrt{\tilde{\varpi}\tilde{\ell}+ \tilde{\kappa} c_1} c_2}{2 \sqrt{ \tilde{\kappa}}}|\frac{2 \tilde{\varpi}\tilde{\ell}}{\tilde{\varpi}\tilde{\ell}+ \tilde{\kappa} c_1}\right)|\frac{2 \tilde{\varpi}\tilde{\ell}}{\tilde{\varpi}\tilde{\ell}+ \tilde{\kappa} c_1}\right)\right)~~~~~~~~~~~~~\nonumber\\
\times \text{cn}\left(\frac{\sqrt{\tilde{\varpi}\tilde{\ell}+ \tilde{\kappa} c_1} \left(c_2+2 \pi \right)}{2 \sqrt{ \tilde{\kappa}}}|\frac{2 \tilde{\varpi}\tilde{\ell}}{\tilde{\varpi}\tilde{\ell}+ \tilde{\kappa} c_1}\right)^2 \text{dn}\left(\frac{\sqrt{\tilde{\varpi}\tilde{\ell}+ \tilde{\kappa} c_1} \left(c_2+2 \pi \right)}{2 \sqrt{ \tilde{\kappa}}}|\frac{2 \tilde{\varpi}\tilde{\ell}}{\tilde{\varpi}\tilde{\ell}+ \tilde{\kappa} c_1}\right)~~~~~~~~~~~~\nonumber\\
\times \sqrt{\frac{-2 \tilde{\varpi}\tilde{\ell} \text{sn}\left(\frac{\sqrt{\tilde{\varpi}\tilde{\ell}+ \tilde{\kappa} c_1} c_2}{2 \sqrt{ \tilde{\kappa}}}|\frac{2 \tilde{\varpi}\tilde{\ell}}{\tilde{\varpi}\tilde{\ell}+ \tilde{\kappa} c_1}\right)^2+c_1  \tilde{\kappa}+\tilde{\varpi}\tilde{\ell}}{c_1  \tilde{\kappa}+\tilde{\varpi}\tilde{\ell}}}~~~~~~~~~~~~~~~~~~~~~~~~\nonumber\\
+c_1 c_2 \sqrt{ \tilde{\kappa}} \sqrt{c_1  \tilde{\kappa}+\tilde{\varpi}\tilde{\ell}}-2 \left(c_1  \tilde{\kappa}+\tilde{\varpi}\tilde{\ell}\right) E\left(\text{am}\left(\frac{\sqrt{\tilde{\varpi}\tilde{\ell}+ \tilde{\kappa} c_1} c_2}{2 \sqrt{ \tilde{\kappa}}}|\frac{2 \tilde{\varpi}\tilde{\ell}}{\tilde{\varpi}\tilde{\ell}+ \tilde{\kappa} c_1}\right)|\frac{2 \tilde{\varpi}\tilde{\ell}}{\tilde{\varpi}\tilde{\ell}+ \tilde{\kappa} c_1}\right)~~~~~~~~~~~\nonumber\\
\times \text{dn}\left(\frac{\left(c_2+2 \pi \right) \sqrt{c_1  \tilde{\kappa}+\tilde{\varpi}\tilde{\ell}}}{2 \sqrt{ \tilde{\kappa}}}|\frac{2 \tilde{\varpi}\tilde{\ell}}{c_1  \tilde{\kappa}+\tilde{\varpi}\tilde{\ell}}\right)~~~~~~~~~~~~~~~~~~~~~~~~~~~~~~~~~~\nonumber\\
\times \left(\text{cn}\left(\frac{\sqrt{\tilde{\varpi}\tilde{\ell}+ \tilde{\kappa} c_1} c_2}{2 \sqrt{ \tilde{\kappa}}}|\frac{2 \tilde{\varpi}\tilde{\ell}}{\tilde{\varpi}\tilde{\ell}+ \tilde{\kappa} c_1}\right)^2-\text{sn}\left(\frac{\sqrt{\tilde{\varpi}\tilde{\ell}+ \tilde{\kappa} c_1} c_2}{2 \sqrt{ \tilde{\kappa}}}|\frac{2 \tilde{\varpi}\tilde{\ell}}{\tilde{\varpi}\tilde{\ell}+ \tilde{\kappa} c_1}\right)^2\right)~~~~~~~~~~~~~~~~~~~~~~~~~\nonumber\\
\times \sqrt{\frac{-2 \tilde{\varpi}\tilde{\ell} \text{sn}\left(\frac{\sqrt{\tilde{\varpi}\tilde{\ell}+ \tilde{\kappa} c_1} c_2}{2 \sqrt{ \tilde{\kappa}}}|\frac{2 \tilde{\varpi}\tilde{\ell}}{\tilde{\varpi}\tilde{\ell}+ \tilde{\kappa} c_1}\right)^2+c_1  \tilde{\kappa}+\tilde{\varpi}\tilde{\ell}}{c_1  \tilde{\kappa}+\tilde{\varpi}\tilde{\ell}}} \text{sn}\left(\frac{\sqrt{\tilde{\varpi}\tilde{\ell}+ \tilde{\kappa} c_1} \left(c_2+2 \pi \right)}{2 \sqrt{ \tilde{\kappa}}}|\frac{2 \tilde{\varpi}\tilde{\ell}}{\tilde{\varpi}\tilde{\ell}+ \tilde{\kappa} c_1}\right)^2~~~~~~~~~~~~~~\nonumber\\
+\left( c_1 \left(c_2+2 \pi \right) \sqrt{ \tilde{\kappa}} \sqrt{c_1  \tilde{\kappa}+\tilde{\varpi}\tilde{\ell}}-2 \left(c_1  \tilde{\kappa}+\tilde{\varpi}\tilde{\ell}\right) E\left(\text{am}\left(\frac{\sqrt{\tilde{\varpi}\tilde{\ell}+ \tilde{\kappa} c_1} \left(c_2+2 \pi \right)}{2 \sqrt{ \tilde{\kappa}}}|\frac{2 \tilde{\varpi}\tilde{\ell}}{\tilde{\varpi}\tilde{\ell}+ \tilde{\kappa} c_1}\right)|\frac{2 \tilde{\varpi}\tilde{\ell}}{\tilde{\varpi}\tilde{\ell}+ \tilde{\kappa} c_1}\right)\right) ~~~\nonumber\\
\times \text{dn}\left(\frac{c_2 \sqrt{c_1  \tilde{\kappa}+\tilde{\varpi}\tilde{\ell}}}{2 \sqrt{ \tilde{\kappa}}}|\frac{2 \tilde{\varpi}\tilde{\ell}}{c_1  \tilde{\kappa}+\tilde{\varpi}\tilde{\ell}}\right)
\left(  \text{cn}\left(\frac{\sqrt{\tilde{\varpi}\tilde{\ell}+ \tilde{\kappa} c_1} c_2}{2 \sqrt{ \tilde{\kappa}}}|\frac{2 \tilde{\varpi}\tilde{\ell}}{\tilde{\varpi}\tilde{\ell}+ \tilde{\kappa} c_1}\right)^2-\text{sn}\left(\frac{\sqrt{\tilde{\varpi}\tilde{\ell}+ \tilde{\kappa} c_1} c_2}{2 \sqrt{ \tilde{\kappa}}}|\frac{2 \tilde{\varpi}\tilde{\ell}}{\tilde{\varpi}\tilde{\ell}+ \tilde{\kappa} c_1}\right)^2\right)~~~~~~~ \nonumber\\
\times \left( \text{cn}\left(\frac{\sqrt{\tilde{\varpi}\tilde{\ell}+ \tilde{\kappa} c_1} \left(c_2+2 \pi \right)}{2 \sqrt{ \tilde{\kappa}}}|\frac{2 \tilde{\varpi}\tilde{\ell}}{\tilde{\varpi}\tilde{\ell}+ \tilde{\kappa} c_1}\right)^2-\text{sn}\left(\frac{\sqrt{\tilde{\varpi}\tilde{\ell}+ \tilde{\kappa} c_1} \left(c_2+2 \pi \right)}{2 \sqrt{ \tilde{\kappa}}}|\frac{2 \tilde{\varpi}\tilde{\ell}}{\tilde{\varpi}\tilde{\ell}+ \tilde{\kappa} c_1}\right)^2\right) ~~~~~~~~~~~~~~~~~~\nonumber\\
\times \sqrt{\frac{-2 \tilde{\varpi}\tilde{\ell} \text{sn}\left(\frac{\sqrt{\tilde{\varpi}\tilde{\ell}+ \tilde{\kappa} c_1} \left(c_2+2 \pi \right)}{2 \sqrt{ \tilde{\kappa}}}|\frac{2 \tilde{\varpi}\tilde{\ell}}{\tilde{\varpi}\tilde{\ell}+ \tilde{\kappa} c_1}\right)^2+c_1  \tilde{\kappa}+\tilde{\varpi}\tilde{\ell}}{c_1  \tilde{\kappa}+\tilde{\varpi}\tilde{\ell}}}~~~~~~~~~~~~~~~~~~~~~~~~~~~~~
\end{eqnarray*}
and,
\begin{eqnarray*}
\mathtt{D}=\tilde{\varpi}\tilde{\ell}\sqrt{c_1 \tilde{\kappa}+\tilde{\varpi}\tilde{\ell}} \text{dn}\left(\frac{\sqrt{\tilde{\varpi}\tilde{\ell}+ \tilde{\kappa} c_1} c_2}{2 \sqrt{ \tilde{\kappa}}}|\frac{2 \tilde{\varpi}\tilde{\ell}}{\tilde{\varpi}\tilde{\ell}+ \tilde{\kappa} c_1}\right) \text{dn}\left(\frac{\sqrt{\tilde{\varpi}\tilde{\ell}+ \tilde{\kappa} c_1} \left(c_2+2 \pi \right)}{2 \sqrt{ \tilde{\kappa}}}|\frac{2 \tilde{\varpi}\tilde{\ell}}{\tilde{\varpi}\tilde{\ell}+ \tilde{\kappa} c_1}\right)\nonumber\\
\times \left( \text{cn}\left(\frac{\sqrt{\tilde{\varpi}\tilde{\ell}+ \tilde{\kappa} c_1} c_2}{2 \sqrt{ \tilde{\kappa}}}|\frac{2\tilde{\varpi}\tilde{\ell}}{\tilde{\varpi}\tilde{\ell}+ \tilde{\kappa} c_1}\right)^2-\text{sn}\left(\frac{\sqrt{\tilde{\varpi}\tilde{\ell}+ \tilde{\kappa} c_1} c_2}{2 \sqrt{ \tilde{\kappa}}}|\frac{2\tilde{\varpi}\tilde{\ell}}{\tilde{\varpi}\tilde{\ell}+ \tilde{\kappa} c_1}\right)^2\right) \nonumber\\
\times\left( \text{cn}\left(\frac{\sqrt{\tilde{\varpi}\tilde{\ell}+ \tilde{\kappa} c_1} \left(c_2+2 \pi \right)}{2 \sqrt{ \tilde{\kappa}}}|\frac{2 \tilde{\varpi}\tilde{\ell}}{\tilde{\varpi}\tilde{\ell}+ \tilde{\kappa} c_1}\right)^2-\text{sn}\left(\frac{\sqrt{\tilde{\varpi}\tilde{\ell}+ \tilde{\kappa} c_1} \left(c_2+2 \pi \right)}{2 \sqrt{ \tilde{\kappa}}}|\frac{2 \tilde{\varpi}\tilde{\ell}}{\tilde{\varpi}\tilde{\ell}+ \tilde{\kappa} c_1}\right)^2\right).
\end{eqnarray*}


\end{document}